\documentclass[11pt]{article}
\usepackage{amsfonts,amsmath,amssymb}
\usepackage{epsfig}
\usepackage{listings}
\usepackage[small,it]{caption} 
\usepackage{verbatim}
\usepackage{todonotes}

\usepackage{color}

\definecolor{dkgreen}{rgb}{0,0.6,0}
\definecolor{gray}{rgb}{0.5,0.5,0.5}
\definecolor{mauve}{rgb}{0.58,0,0.82}

\newlength{\mylength}
\usepackage{amsthm}

\theoremstyle{definition}

\title{Compiling a Q\# Subset to QASM 3.0 in TypeScript via a JSON Based IR}

\author{Marcus Edwards \\ Electrical and Computer Engineering, \\University of British Columbia, Vancouver, BC, CA}

\begin{document}

\maketitle

\begin{abstract}
We implement a compile toolchain from Q\# to QASM 3.0 including a full-featured lexer and parser implementation, as well as 
a compiler that supports a subset of Q\# features. The lexer, parser and compiler are shown to work with various input Q\# programs
and the implementation is compared against existing Q\# compile tools. Unlike the Microsoft implementation of the official Q\# compile toolchain,
our implementation is written in TypeScript in order to port functionality to web environments.
\end{abstract}

\section{Introduction}

Microsoft's Q\# \cite{Svore_2018} offers a strongly typed system of high-level abstractions on top of 
your typical quantum instruction set architecture including classical and quantum operations.
In particular, Q\# includes a strong type system, contextual expressions, callables, iterables, conditional branching, 
loops, literals, closures, quantum memory management tools similar to those offered in Rust (i.e. use and borrow statements),
namespaces, comments as well as features specific to certain quantum computing techniques such as conjugations and operation modifiers (i.e. Controlled and Adjoint).
We implement a full-featured parser of Q\# language syntax into custom Abstract Syntax Tree (AST)
data structure which takes some inspiration from existing npm libraries including Qasm-ts \cite{kim2024enablingverificationformalizationhybrid}, 
Blackbird-ts and Qmasm-ts \cite{edwards2023quantumprogramminglanguageparser}. Q\# represents the most 
complex syntax to parse out of these four, and therefore required a more complex parser implementation.
We also implement a compiler of a subset of Q\# language features to QASM 3.0 syntax \cite{Cross_2022}.
Our implementation is object oriented and is based on TypeScript, which has been shown previously to be 
a reasonable ``middle ground'' option which is more efficient than Python for similar applications, though less so than Rust, which 
can only be executed in more restrictive contexts \cite{kim2024enablingverificationformalizationhybrid}.
The pieces of the compile toolchain can be used independently and we will cover the details of their implementations 
in the following sections.

\section{Lexer Implementation}

The toolchain begins with the lexer. The job of the lexer is to translate the input Q\# code to a sequence of meaningful tokens.
These tokens represent meaningful collections of characters that correspond to the atomic elements of the Q\# syntax. These include 
operators such as bitwise operators !!!, \&\&\&, for example, which are meaningful when considered together but needlessly 
complicated to interpret if left as a sequence of three elements. The lexer is not responsible for generating all the idiomatic abstractions
we are interested in, however. So, an expression composed of bitwise operators is transformed into a sequence of bitwise operator tokens like 
`Token.BitwiseAnd' and `Token.BitwiseNot' rather than into any abstraction like an `Expression' which would have its own elements and an abstraction
depth greater than one. Abstractions of any depth greater than one are left to the parser, whose job it is to generate these higher-level abstractions from the tokens 
that the lexer first generates.

The lexer has some concept of what categories the Tokens fit into, i.e. whether they may be parts of `Expressions' or `Parameters', or not, 
and provides some conveniece functions for checking these things. These are a convenience to the parser.

The set of supported tokens is derived from the syntax of Q\# which is detailed in Microsoft's online documentation. The lexer also knowns Q\#'s 
keywords. It is therefore very good at inferring when a series of characters is a keyword that should be represented by its own individual `Token', 
a `String' literal, or an `Identifer', for example.

The lexer is the least complex part of the compile toolchain, while the parser is the most complex.

\section{Parser Implementation}

The parser implementation is the most complex part of the compile toolchain presented. It uses recursive descent parsing to traverse an entire program.
AST nodes are built up one at a time while each line of code is parsed by a call to `parseNode()'. In some cases, a new parser is instantiated for a sub context 
such as a function or an operation which should have its own namespace. This allows the parser to infer when references are made to variables versus other identifiers
properly, considering the scope of any declarations. In this case, the parser advances the cursor through the token sequence until the correct closing bracket is reached, demarking the 
end of the scope parsed by the sub context parser.

The `parseNode' method is responsible for delegating the parsing of a given construct to the appropriate parsing method, such as `twoQubitGate()', `isingRotationGate()' 
or `conjugation()' among many others, which each accept the sequence of remaining tokens as a parameter. In each of these methods, one or more AST nodes are instantiated and returned 
so that they may be appended to the AST at the appropriate level. Any information pertaining to a node, such as the rotation angle, target and control qubits involved in a two qubit gate like `Rxx', are 
pulled out of the token sequence systematically and used to instantiate the `Rxx' AstNode. In many cases, nodes themselves have attributes which are nodes or sequences of nodes, such as a `While' loop 
that contains a block of code which must be parsed itself, and a condition which must also be parsed into an `Expression' node.

`Expressions' are the most flexible type of node in the AST and involve free-formed sequences of `Parameters' which each have a representation used to recursively build the representations of higher order `Expressions'.
These expressions can involve `Variables', `Identifiers', `Ints', `Strings', indexes including `Ints', `Ranges' and other sub-expressions among nearly all of the types in the Q\# types system. However, declaring 
`Functions' and `Operations' or providing `For' or `While' loops, for example, within an expression is (obviously) not allowed. `Functions', etc. should be referred to by their names.
`Expressions' are found everywhere in Q\# such as in `Function' parameters, the right hand sides of `Variable' declarations, and so on.

Keeping track of `Expression' representations as strings makes compiling them easier, since the compiler can refer to any `Expression's representation in the compiler pass, for example 
generating the matching QASM code for a complex expression with a succinct reference to `node.repr' which is already prepared and accounts for any recursive inclusion of sub `Expressions'
and `Parameters' of virtually any type. The only thing that is needed to compile an expresion is to replace those strings that are used to represent certain operators and idioms in Q\# to the 
corresponding ones in QASM. This is done in the compiler.

The parser is able to construct an AST with the following features: clauses, single qubit gates, two qubit gates, single qubit gates involving rotations, arbitrary unitaries given by complex matrices, 
two qubit gates involving rotations, free-form `Expressions', `Conjugations', `Variables', `Qubits', `Operations', fail statements, `Modifiers', `Functions', symbols, `Identifiers', applications of non-intrinsic `Operators',
lists of indices, lists of `Parameters', `Parameters', `Structs', `Arrays', `Use' statements, `Borrow' statements, `Imports', `Variable' declarations and assignments, `Mutable' declarations and assignments, `Conditions', Q\# types,
`Ranges', for loops, while loops, repeat loops, lists of `Ints', lists of symbols, `Ints', and more. 

For example, the folloring AST structure results when parsing this segement of the Grover's algorithm code in Q\#. Here each Object is printed in JSON which does not include the 
Object's type. However, the type information is indicated in comments we have provided and is available to the compile at compile time.

\vspace{5mm}
\noindent
Q\# Operation:
\begin{lstlisting}
operation ReflectAboutMarked(inputQubits : Qubit[]) : Unit {
    use outputQubit = Qubit();
    within {
        // We initialize the outputQubit to (|0> - |1>) / sqrt(2), so that
        // toggling it results in a (-1) phase.
        X(outputQubit);
        H(outputQubit);
        // Flip the outputQubit for marked states.
        // Here, we get the state with alternating 0s and 1s by using the X
        // operation on every other qubit.
        for q in inputQubits[...2...] {
            X(q);
        }
    } apply {
        Controlled X(inputQubits, outputQubit);
    }
}
\end{lstlisting}

\noindent
Parsed Abstract Syntax Tree Segment (we have added comments):
\begin{lstlisting}
// operation
{
    "name": "ReflectAboutMarked",
    "nodes": [
        // qubit allocation
        {
            "name": {
                "repr": "outputQubit",
                "val": "outputQubit"
            },
            "qubits": {
                "repr": "outputQubit",
                "name": "outputQubit",
                "length": {
                    "repr": "1",
                    "val": 1
                }
            }
        },
        // within / apply closure
        {
            "within": [
                // comment
                {
                    "val": " We initialize the outputQubit to (|0> - |1>) / sqrt(2), so that"
                },
                // comment
                {
                    "val": " toggling it results in a (-1) phase."
                },
                // Pauli X gate
                {
                    "target": {
                        "repr": "outputQubit",
                        "id": "outputQubit"
                    }
                },
                // Hadamard gate
                {
                    "target": {
                        "repr": "outputQubit",
                        "id": "outputQubit"
                    }
                },
                // comment
                {
                    "val": " Flip the outputQubit for marked states."
                },
                // comment
                {
                    "val": " Here, we get the state with alternating 0s and 1s by using the X"
                },
                // comment
                {
                    "val": " operation on every other qubit."
                },
                // for loop
                {
                    "variable": {
                        "repr": "q",
                        "name": "q"
                    },
                    "inside": [
                        // Pauli X gate
                        {
                            "target": {
                                "repr": "q",
                                "id": "q"
                            }
                        }
                    ],
                    "vals": {
                        "repr": "inputQubits[...2...]",
                        "vals": [
                            // expression for loop iterable
                            {
                                "repr": "inputQubits[...2...]",
                                "instance": "inputQubits",
                                "index": {
                                    "repr": "...2...",
                                    "lower": {},
                                    "upper": {}
                                }
                            }
                        ],
                        "size": 1
                    }
                }
            ],
            "applies": [
                // controlled X gate
                {
                    "control": {
                        "repr": "inputQubits",
                        "id": "inputQubits"
                    },
                    "target": {
                        "repr": "outputQubit",
                        "id": "outputQubit"
                    }
                }
            ]
        }
    ],
    "params": [
        [
            // input parameter expression
            {
                "repr": "inputQubits",
                "elements": [
                    {
                        "repr": "inputQubits",
                        "id": "inputQubits"
                    }
                ]
            }
        ]
    ],
    // the operation has no modifiers
    "modifiers": [],
    // the operation does not return anything
    "returnType": {}
}
\end{lstlisting}

To demonstrate how the `Expressions' are parsed, we also include the following example from the same Grover.qs file.

\vspace{5mm}
\noindent
\begin{lstlisting}
function CalculateOptimalIterations(nQubits : Int) : Int {
    if nQubits > 63 {
        fail "This sample supports at most 63 qubits.";
    }
    let nItems = 1 <<< nQubits; // 2^nQubits
    let angle = ArcSin(1. / Sqrt(IntAsDouble(nItems)));
    let iterations = Round(0.25 * PI() / angle - 0.5);
    return iterations;
}

\end{lstlisting}

\noindent
Parsed Function:
\begin{lstlisting}
    // function
    {
        "name": "CalculateOptimalIterations",
        "nodes": [
          // if
          {
            "condition": 
            // condition expression
            {
              "repr": "nQubits>63",
              "elements": [
                // qubit identifier
                {
                  "repr": "nQubits",
                  "id": "nQubits"
                },
                // greater than operator
                {
                  "repr": ">"
                },
                // integer literal
                {
                  "repr": "63",
                  "val": 63
                }
              ]
            },
            "ifClause": [
                // fail statement
              {
                "msg": {
                  "repr": [
                    51,
                    "\"This sample supports at most 63 qubits.\""
                  ],
                  "val": [
                    51,
                    "\"This sample supports at most 63 qubits.\""
                  ]
                }
              }
            ]
          },
          // let variable declaration and assignment
          {
            "expression": 
            // right hand side expression
            {
              "repr": "1<<<nQubits",
              "elements": [
                // literal
                {
                  "repr": "1",
                  "val": 1
                },
                // bitwise operator
                {
                  "repr": "<<<"
                },
                // qubit identifier
                {
                  "repr": "nQubits",
                  "id": "nQubits"
                }
              ]
            },
            "variable": {
              "repr": "nItems",
              "name": "nItems"
            }
          },
          // let variable declaration and assignment
          {
            "expression": 
            // right hand side expression
            {
              "repr": "ArcSin(1/Sqrt(IntAsDouble(nItems, ), ), )",
              "elements": [
                // ArcSin function call
                {
                  "repr": "ArcSin(1/Sqrt(IntAsDouble(nItems, ), ), )",
                  "name": "ArcSin",
                  "params": [
                    [
                      // ArcSin parameter expression
                      {
                        "repr": "1/Sqrt(IntAsDouble(nItems, ), )",
                        "elements": [
                          // Double literal
                          {
                            "repr": "1",
                            "val": 1
                          },
                          // divide by operator
                          {
                            "repr": "/"
                          },
                          // Sqrt function call
                          {
                            "repr": "Sqrt(IntAsDouble(nItems, ), )",
                            "name": "Sqrt",
                            "params": [
                              [
                                // Sqrt function parameter expression
                                {
                                  "repr": "IntAsDouble(nItems, )",
                                  "elements": [
                                    //  IntAsDouble function call
                                    {
                                      "repr": "IntAsDouble(nItems, )",
                                      "name": "IntAsDouble",
                                      "params": [
                                        [
                                          // IntAsDouble function param expression
                                          {
                                            "repr": "nItems",
                                            "elements": [
                                              // variable reference
                                              {
                                                "repr": "nItems",
                                                "name": "nItems"
                                              }
                                            ]
                                          }
                                        ]
                                      ]
                                    }
                                  ]
                                }
                              ]
                            ]
                          }
                        ]
                      }
                    ]
                  ]
                }
              ]
            },
            "variable": {
              "repr": "angle",
              "name": "angle"
            }
          },
          {
            "expression": 
            // let variable declaration and assignment
            {
              "repr": "Round(0.25*PI()/angle-0.5, )",
              "elements": [
                // right hand side expression
                {
                  "repr": "Round(0.25*PI()/angle-0.5, )",
                  "name": "Round",
                  "params": [
                    [
                      // Round function parameter expression
                      {
                        "repr": "0.25*PI()/angle-0.5",
                        "elements": [
                          // Double literal
                          {
                            "repr": "0.25",
                            "val": 0.25
                          },
                          // multiplier operator
                          {
                            "repr": "*"
                          },
                          // PI function call
                          {
                            "repr": "PI()",
                            "name": "PI",
                            "params": []
                          },
                          // divide operator
                          {
                            "repr": "/"
                          },
                          // variable reference
                          {
                            "repr": "angle",
                            "name": "angle"
                          },
                          // minus operator
                          {
                            "repr": "-"
                          },
                          // Double literal
                          {
                            "repr": "0.5",
                            "val": 0.5
                          }
                        ]
                      }
                    ]
                  ]
                }
              ]
            },
            "variable": {
              "repr": "iterations",
              "name": "iterations"
            }
          },
          // return statement 
          {
            "expr": {
              "repr": "iterations",
              "elements": [
                // return statement expression
                {
                  "repr": "iterations",
                  "name": "iterations"
                }
              ]
            }
          }
        ],
        "params": [
          [
            // function parameters expression
            {
              "repr": "nQubits",
              "elements": [
                {
                  "repr": "nQubits",
                  "id": "nQubits"
                }
              ]
            }
          ]
        ]
      }
\end{lstlisting}

\section{Compiler Implementation}

The compiler implementation checks for any scopes such as loops, functions and operations that are fully supported by 
the compiler and compiles these into corresponding QASM 3.0 functions, loops, etc. The compilation makes full use of the types 
and attributes of objects in the AST to generate corresponding QASM code. In the end, this means translating Q\# syntax to QASM syntax.
It also involves in some cases translating intrinsic Q\# instructions, which are often more high-level, to the lower-level 
instruction set provided by QASM 3.0. An example of this is given by the translation of two-qubit Ising XX, YY and ZZ rotations, 
which are intrinsic operators in Q\#, to collections of quantum instructions in QASM.

\vspace{5mm}
\noindent
Input Q\#:
\begin{lstlisting}
use register = Qubit[2];

// Ising XX 
Rxx(1.5, register[0], register[1]);

// Ising XX
Rxx(2.3, register[0], register[1]);

// Ising ZZ
Rzz(8.5, register[0], register[1]);
\end{lstlisting}

\noindent
Output QASM 3.0:
\begin{lstlisting}
qubit[2] register;

//  Ising XX 
u3(pi/2, 1.5, 0) register[0];
h register[1];
cx register[0],register[1];
u1(-1.5) register[1];
cx register[0],register[1];
h register[1];
u2(-pi, pi-1.5) register[0];

//  Ising XX
u3(pi/2, 2.3, 0) register[0];
h register[1];
cx register[0],register[1];
u1(-2.3) register[1];
cx register[0],register[1];
h register[1];
u2(-pi, pi-2.3) register[0];

//  Ising ZZ
cx register[0],register[1];
u1(8.5) register[1];
cx register[0],register[1];
\end{lstlisting}

\noindent
Compiler code:
\begin{lstlisting}
if (node instanceof Rxx) {
    node.qasmString = `u3(pi/2, ${node.rads.repr}, 0) ${node.qubit0.repr};\nh ${node.qubit1.repr};\ncx ${node.qubit0.repr},${node.qubit1.repr};\nu1(-${node.rads.repr}) ${node.qubit1.repr};\ncx ${node.qubit0.repr},${node.qubit1.repr};\nh ${node.qubit1.repr};\nu2(-pi, pi-${node.rads.repr}) ${node.qubit0.repr};\n`;
} else if (node instanceof Ryy) {
    node.qasmString = `cy ${node.qubit0.repr},${node.qubit1.repr};ry(${node.rads.repr}) ${node.qubit0.repr};cy ${node.qubit0.repr},${node.qubit1.repr};\n`
} else if (node instanceof Rzz) {
    node.qasmString = `cx ${node.qubit0.repr},${node.qubit1.repr};\nu1(${node.rads.repr}) ${node.qubit1.repr};\ncx ${node.qubit0.repr},${node.qubit1.repr};\n`;
}
\end{lstlisting}

These and more examples of compilation can be found in `demo.qs' and `demo.qasm' in the `spec' directory. The lexer, parser, compiler toolchain 
can also be run against other inputs using either the functions exported from `main.ts' or by running `example.ts' which is meant for this purpose!

In certain specific cases, type information can be inferred at compile time. This is however not always possible since a robust import system was not implemented 
due to time constraints. Therefore, some imported variables have indeterminate types, for example. When a literal is provided directly however, it is easy to infer type 
information and this is used to optimize the generated QASM 3.0 code i.e.

\vspace{5mm}
\noindent
Inferring QASM types for a loop:
\begin{lstlisting}
if (ty == 'Int' || ty == 'BigInt') {
    if (Math.abs(max) < 128 ) {
        scopeQasm += 'int[8] '
    } else if (Math.abs(max) < 32768) {
        scopeQasm += 'int[16] '
    } else if (Math.abs(max) < 2147483648) {
        scopeQasm += 'int[32] '
    } else if (Math.abs(max) < 9223372036854775808) {
        scopeQasm += 'int[64] '
    } else {
        scopeQasm += 'int '  // QASM does not have BigInt
    }
} else if (ty == 'Double') {
    scopeQasm += 'float '
} else if (ty == 'Bool') {
    scopeQasm += 'bool '
} else if (ty == 'Qubit') {
    scopeQasm += 'qubit '
}
\end{lstlisting}

Here the variable `ty' refers to the type of the node in the abstract syntax tree currently being compiled.
`Identifier', `Variable' and `GetParam' are also valid types of AST nodes whose corresponding Q\# types are not inferred (in 
the current implementation, for now) due to the complexity of inferring the types of imported variables and libraries
which are commonplace in Q\# programs. A secondary complexity which prohibits us from implementing this level of depth in 
our type inference is that variables are not evaluated at compile time. Therefore, there would be no way for us to tell if 
a Variable of `IntType' were too large to be sompiled to a `int[8]' QASM variable which might be used in various contexts.
We can only optimize therefore when we have a literal: an `Int' or `BigInt' instead of a `Variable' of `IntType'.

Like the parser, the compiler uses a recursive descent approach to traversing the input data structure, which is the AST in this case.
Among the supported Q\# features are comments, functions, operations, repeat clauses, while loops, for loops, use statements and conditionals. The entire intrinsic 
gate library is also supported including X, Y, Z, T, SWAP, S, Rxx, Ryy, Rzz, Rx, Ry, Rz, ResetAll, Reset, R1, R1Frac, RFrac, R, Measure,
M, I, H, CNOT, CZ, CCNOT. Many of these are not included in the QASM instruction set architecture and must be compiled to sets of QASM instructions.

\section{Discussion}

Several challenges were encountered in the implementation of this package. The first challenge was simply that the Q\# language has  very many features and idioms as 
compared to the lower level assembly language style languages we have implemented similar tools for in the past. While QASM 3.0 does introduce many classical control 
flow constructs that are similar to Q\# language features, QASM 2.0 did not have these. Therefore, despite the complexity of the Q\# and QASM 3.0 languages causing some headache
at first, choosing to compile to QASM 3.0 which has these similarities in the end made the compiler pass a bit simpler. This pairing was sort of a two-edged sword.

The second challenge was that many of the Q\# intrinsic operators were more abstract and higher level even than those provided in QASM 3.0. For example, consider how the `Measure'
operator of Q\# is compiled to QASM 3.0.

\begin{lstlisting}
if (node instanceof Measure) {
    let qasmString = '';
    if (node.basis.val == Paulis.PauliX) {
        for (let qubit of node.qubits) {
            qasmString += `h ${qubit.repr};\n`;  // put each qubit in the X basis
        }
    } else if (node.basis.val == Paulis.PauliY) {
        for (let qubit of node.qubits) {
            qasmString += `sdg ${qubit.repr};\n`;  // put each qubit in the Y basis
            qasmString += `h ${qubit.repr};\n`;
        }
    } else if (node.basis.val != Paulis.PauliZ) {
        throw BadArgumentError;
    }
    for (let qubit of node.qubits) {
        qasmString += `measure ${qubit.repr} -> c[${this.qubits.indexOf(qubit)}];\n`;
    }
    node.qasmString = qasmString;
}
\end{lstlisting}

The Q\# `Measure' operator accepts a basis whereas the QASM version of this operation measures in the Z basis.
Therefore, additional operators need to be inserted to transform the subit into the appropriate basis before measuring in 
a QASM program. This is one more example of how Q\# remains higher level in many regards than QASM 3.0, even considering some
gate libraries that extend the QASM standrard library such as `qelib1.inc' which still does not contain all the gates in Q\#'s 
intrinsic library. This makes us confident that our technology can be considered a compiler (which \textit{lowers} Q\# to QASM) rather 
than a transpiler (which would translate between two languages that are siblings in the abstraction hierarchy of the quantum programming stack).

The compile toolchain presented can be compared against the one maintained by Microsoft \cite{Svore_2018}. For example, the compilation of the Grover.qs program 
using our package takes a total time of 7.60 milliseconds. This is as opposed to the 7.29 milliseconds that Microsoft's toolchain takes to compile and 
execute the same program. Our compile time of 7.60 ms can be broken down further into 1.12 ms for the final compilation pass, 4.76 ms for the parsing and 
1.72 ms for the lexing. This division of time matches our expectation since the parsing is the most complex step. We observe that our implementation has a similar 
runtime to Microsoft's despite the Microsoft implementation being primarily written in Rust.

A future direction would be to look at implementing a robust import system. This is not out of the question, as an import system was implemented 
in the case of the Qmasm-ts package, for example. However, the Qmasm language \cite{pakin} is much simpler and has a straightforward Macro system that made this easy.

The fact that a QASM parser has previously been implemented in TypeScript also implies that a compiler could be implemented in an alternative fashion to our current implementation.
This alternative approach would involve transforming `AstNodes' of the sort generated by the Q\# parser to those generated by the QASM parser, thereby transforming an AST to an AST 
rather than directly generating QASM code from the Q\# AST. One could argue that this represents a better separation of concerns, since it would be the task of the QASM package to 
generate QASM from its AST and the Q\# package would only then need to be concerned about Q\# and its workings.

The toolchain also has not been thoroughly tested due to time constraints. Therefore, we expect that there may be bugs that could be found and corrected by writing 
a robust set of unit tests. The lexer, parser and compiler have also only been run against a few input Q\# programs to verify that they basically work as intended.

\section{Conclusion}

We conclude that our first approximation of a TypeScript implementation of a compile toolchain for Q\# to QASM 3.0 manages to port functionality to one of the most portable 
and interoperable environments available in the software ecosystem while incurring little (but not zero) additional cost in terms of efficiency over the Microsoft implementation.

\bibliographystyle{alpha}
\bibliography{biblio}

\end{document}